\documentclass[a4paper,11pt]{article}
\usepackage{pos}
\usepackage{graphicx}
\usepackage{lineno}
\def\logo{} 

\title{Model-independent searches and anomaly detection at the CMS experiment}

\ShortTitle{Model-independent searches and anomaly detection at CMS}

\author*[a]{Tamas Almos Vami}

\affiliation[a]{On behalf of the CMS Collaboration\\
  University of California, Santa Barbara, Santa Barbara, CA 93106, USA}

\emailAdd{tamas.almos.vami@cern.ch}

\abstract{The absence of a clear signal of physics beyond the standard model at the CERN LHC motivates search strategies that do not presuppose a specific signal hypothesis. This proceedings present machine-learning-based anomaly detection to search for new physics in a model-agnostic way. The first such search by CMS looks for dijet resonances whose jets have substructure atypical of jets initiated by light quarks or gluons, using five complementary anomaly detection methods applied to 138\,fb$^{-1}$ of proton-proton collision data at $\sqrt{s} = 13$\,TeV. The technique has been validated directly on data by recovering the top quark without the use of labels. The program has also moved into real time: two unsupervised algorithms, AXOL1TL and CICADA, now run on field-programmable gate arrays inside the CMS level-1 trigger and selected more than four billion collision events during 2024 data taking. Finally, the extension of resonant anomaly detection to event-level observables is outlined.}

\FullConference{43rd International Conference on High Energy Physics (ICHEP 2026)\\
30 July  to 5 August , 2026\\
Natal, Brazil\\}

\begin{document}
\maketitle

\section{Introduction}
\label{sec:intro}

More than a decade after the discovery of the Higgs boson, no additional elementary particle has been
observed at the LHC. The space of theories that extend the standard model (SM) is vast, yet the
experimental program has been organized around dedicated searches, each optimized for one specific
hypothesis. Such searches are powerful, but they necessarily leave
gaps: one cannot design an analysis for every conceivable signature, and signals with cross sections
below the reach of inclusive searches remain uncovered. Are we looking for the wrong models, and have
we even thought of the right one?

Anomaly detection offers a complementary strategy: instead of asking whether the data are compatible
with a particular signal, one asks whether they contain events that are unusual with respect to
everything else recorded. Machine learning (ML) makes this practical, because the notion of "unusual"
can be learned from data rather than specified by hand. CMS~\cite{CMS,CMSRun3} is pursuing this idea in three
directions, summarized here: in jet substructure, at the trigger, and at the level of event-wide
observables.

\section{Anomalous jets in the dijet final state}
\label{sec:dijet}

The first ML-based model-agnostic search performed by CMS looks for a narrow heavy resonance \textbf{A}
decaying to two lighter resonances \textbf{B} and \textbf{C}, each of which decays hadronically~\cite{EXO22026}. The
hierarchy $m_\mathrm{B}, m_\mathrm{C} \ll m_\mathrm{A}$ implies that \textbf{B} and \textbf{C} are produced with a
large Lorentz boost, so that their decay products are collected into single large-radius jets. These
jets carry substructure, such as multiple prongs or heavy-flavor content, atypical of the quantum
chromodynamics (QCD) multijet production that dominates the dijet final state, and no assumption is
made on the identity, spin, or couplings of \textbf{B} and \textbf{C}. The analysis uses 138\,fb$^{-1}$ recorded in
2016 to 2018 at $\sqrt{s} = 13$\,TeV. The methods are documented in detail in Ref.~\cite{MLG23002}.

Jets are clustered from particle-flow candidates with the anti-$k_\mathrm{T}$ algorithm using a
distance parameter $R = 0.8$. Events must contain at least two jets with transverse momentum
$p_\mathrm{T} > 300$\,GeV and $\lvert \eta \rvert < 2.5$. The two highest $p_\mathrm{T}$ jets are
taken as the \textbf{B} and \textbf{C} candidates. The dijet invariant mass must satisfy
$m_\mathrm{jj} > 1455$\,GeV so that the triggers are fully efficient, and the pseudorapidity
separation of the two jets must satisfy $\Delta\eta_\mathrm{jj} < 1.3$, which suppresses the
$t$-channel QCD background relative to the $s$-channel signal. Each method then assigns an anomaly
score to every event, and a selection keeps only the 1\% most anomalous events. The dijet mass spectrum of those events is scanned for a localized excess on
top of a smoothly falling background modeled by a parametric fit to the same spectrum, and upper
limits are set. The identical selection without any substructure
requirement defines an inclusive dijet search, which serves as the model-agnostic reference. Five methods spanning three
learning paradigms are deployed in parallel~\cite{MLG23002}.

\subsection{Five complementary anomaly detection methods}
\label{sec:methods}

\textit{VAE-QR} is unsupervised. A variational autoencoder (VAE)~\cite{VAE} is trained on the 100 highest
$p_\mathrm{T}$ constituents of a jet using background-dominated data, so it reconstructs QCD jets
well and anomalous jets poorly. The reconstruction error is the anomaly score. Since rare high-mass
events would otherwise be flagged as anomalous simply for being rare, a quantile regression (QR)
network~\cite{QR} defines the score threshold as a function of $m_\mathrm{jj}$, which preserves the
shape of the background spectrum by construction.

\textit{CWoLa Hunting}~\cite{CWOLAHUNT}, \textit{TNT}~\cite{TNT}, and \textit{CATHODE}~\cite{CATHODE}
are weakly supervised. They follow the classification-without-labels paradigm~\cite{CWOLA}, in which
a classifier separates two mixed samples of data events rather than individually labeled events. One
sample is potentially signal enriched and the other nearly pure background. If a signal is present
the classifier learns to isolate it, and if not, the two samples are indistinguishable and the output
is noise. The methods differ in how the background-like sample is built. CWoLa Hunting takes it from
the sidebands of a sliding $m_\mathrm{jj}$ window. TNT additionally uses an unsupervised autoencoder
applied to one jet to enrich the signal fraction of the sample used to train the classifier for the
other jet, exploiting the fact that in a true signal both jets are anomalous. CATHODE instead learns
the conditional density of the background with a normalizing flow trained outside the signal window,
interpolates it into the window, and samples synthetic background events from it. The
\textit{CATHODE-b} variant adds the b tagging score of each jet to the inputs.

\textit{QUAK}~\cite{QUAK} is semisupervised and sits between the fully model-agnostic methods and a
dedicated search. Normalizing flows trained on simulated signals encode a loose prior on what an
anomaly might look like, and a further flow trained on simulated background provides rejection
power. Events are placed in a two-dimensional space of background-like and signal-like scores, and
the region preferentially populated by signal region events defines the selection.

\subsection{Discovery potential and results}
\label{sec:performance}

\begin{figure}[htbp]
    \centering
    \includegraphics[height=4.8cm]{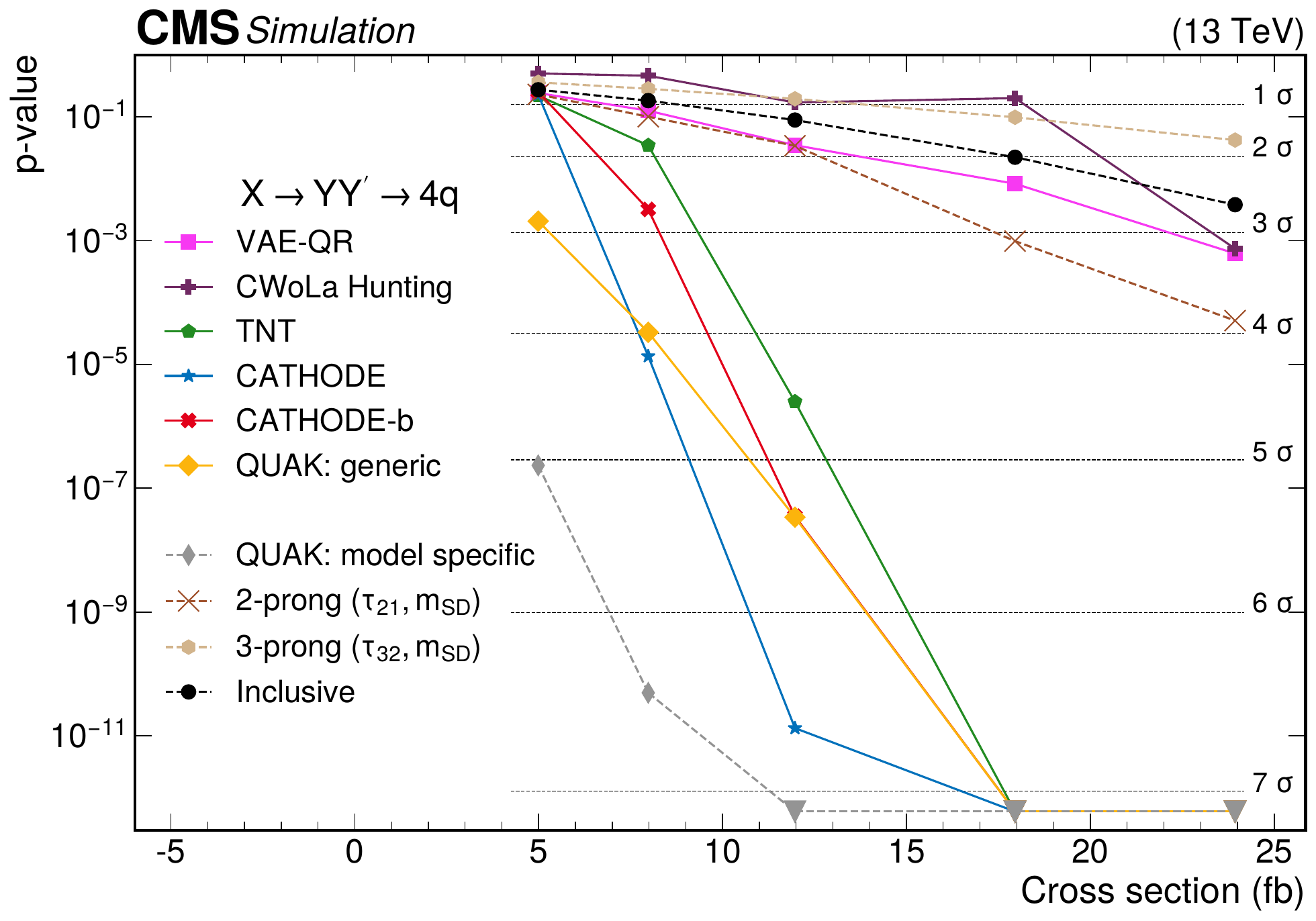}\hfill
    \includegraphics[height=4.8cm]{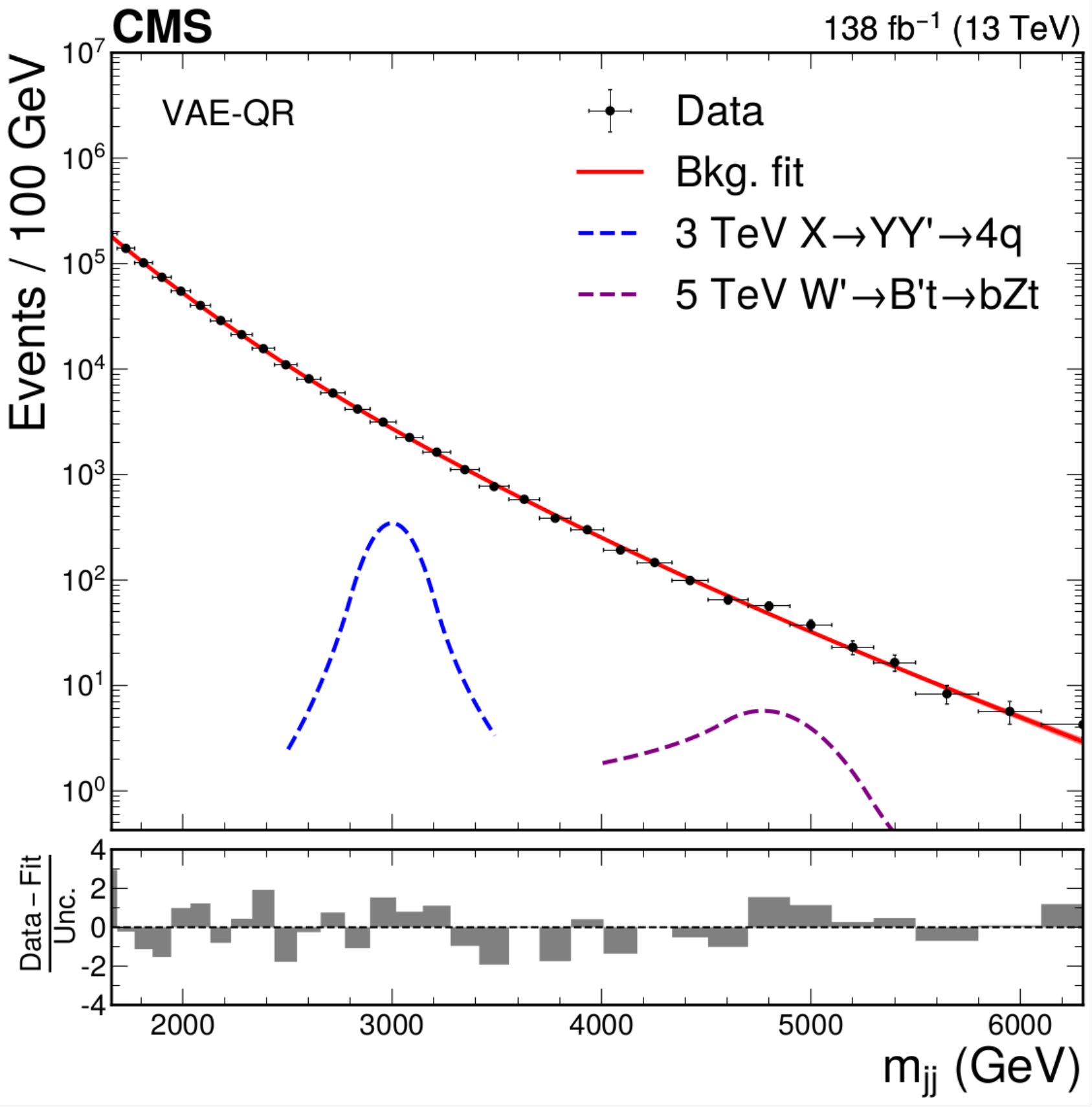}
    \caption{Left: the $p$-value versus injected signal cross section for the
    $\mathrm{X} \to \mathrm{Y}\mathrm{Y}' \to 4\mathrm{q}$ benchmark ($m_\mathrm{X} = 3$\,TeV,
    $m_\mathrm{Y} = m_{\mathrm{Y}'} = 170$\,GeV), comparing the five anomaly detection methods with an
    inclusive dijet search, with 2-prong and 3-prong substructure selections, and with a
    model-specific QUAK configuration. Right: the dijet mass spectrum and background-only fit after
    the VAE-QR selection~\cite{EXO22026}.}
    \label{fig:perf}
\end{figure}

The sensitivity of each method is quantified by injecting signals of varying cross section into a
simulated pseudo-data and repeating the full analysis, including retraining. The resulting
$p$-value as a function of injected cross section is shown on the left of Fig.~\ref{fig:perf} for a
2-prong benchmark. Three features are worth emphasizing. Every anomaly detection method is more
sensitive than the inclusive search, for this signal and for a 3-prong benchmark alike. All of them
nevertheless remain less sensitive than a dedicated search that knows the signal, represented by the
model-specific QUAK configuration: anomaly detection buys generality at a measurable but modest cost.
The traditional substructure selections are no substitute, since the 2-prong tagger improves on the
inclusive search for the 2-prong signal but is markedly worse than doing nothing at all for the
3-prong signal, and vice versa. 

No deviation from the background-only hypothesis with a local significance above $3\sigma$ was found
by any of the five methods. The largest excesses were 2.9, 2.6, 2.3, and $2.2\sigma$ for CATHODE-b,
QUAK, VAE-QR, and CATHODE, at resonance masses of 2.3, 4.7, 4.9, and 2.3\,TeV, respectively, while
TNT and CWoLa Hunting reported nothing above $1.5\sigma$. The fitted dijet mass spectrum for the
VAE-QR selection is shown on the right of Fig.~\ref{fig:perf}. The other methods look similar.
Exclusion limits at 95\% confidence level were set on benchmark models covering 2+2, 3+3, 4+2, 5+5,
and 6+6 prong topologies, and for most of these no dedicated search exists in the mass range
considered, so these are the first limits of their kind. The gain over the inclusive search is
largest for the highest-multiplicity final state, the 6+6 prong
$\mathrm{G}_\mathrm{KK} \to \mathrm{HH} \to 4\mathrm{t}$ model, where the cross section required for
a $5\sigma$ discovery is reduced by a factor of 6.4 and the expected limit improves by up to a factor
of 7.1.

\subsection{Validation on data and interpretation of an excess}
\label{sec:top}

\begin{figure}[htbp]
    \centering
    \includegraphics[height=4.3cm]{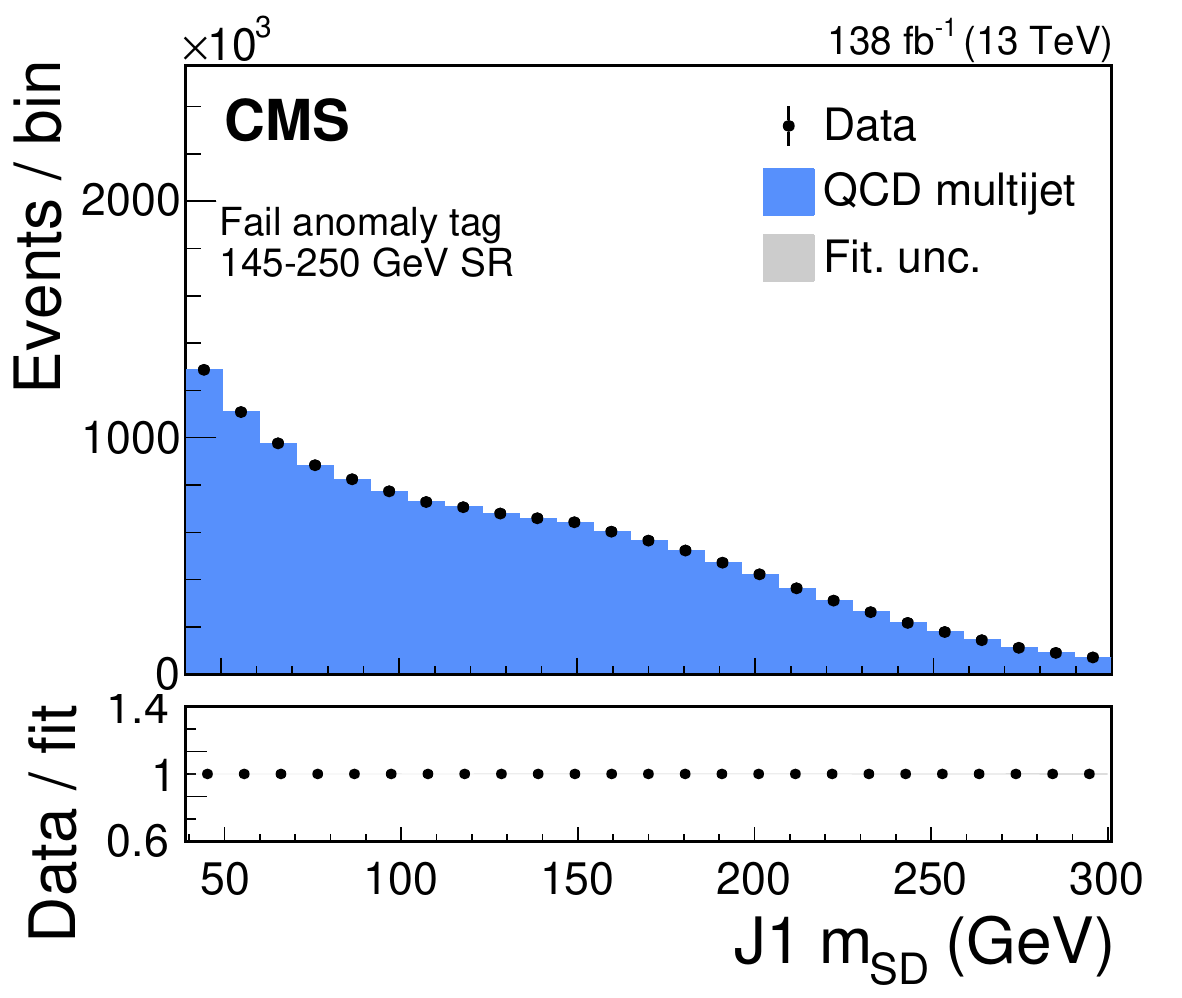}\hfill
    \includegraphics[height=4.3cm]{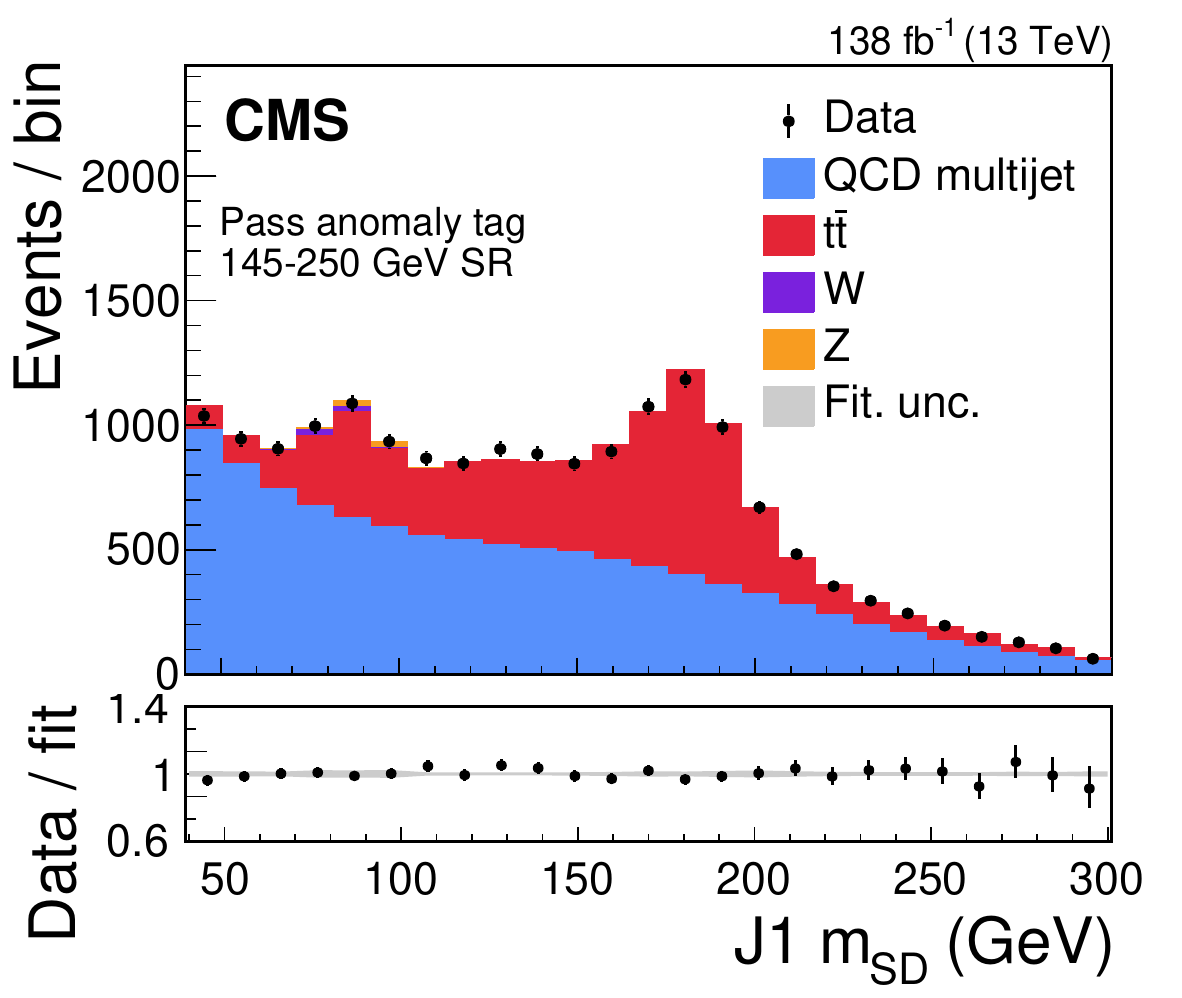}\hfill
    \includegraphics[height=4.3cm]{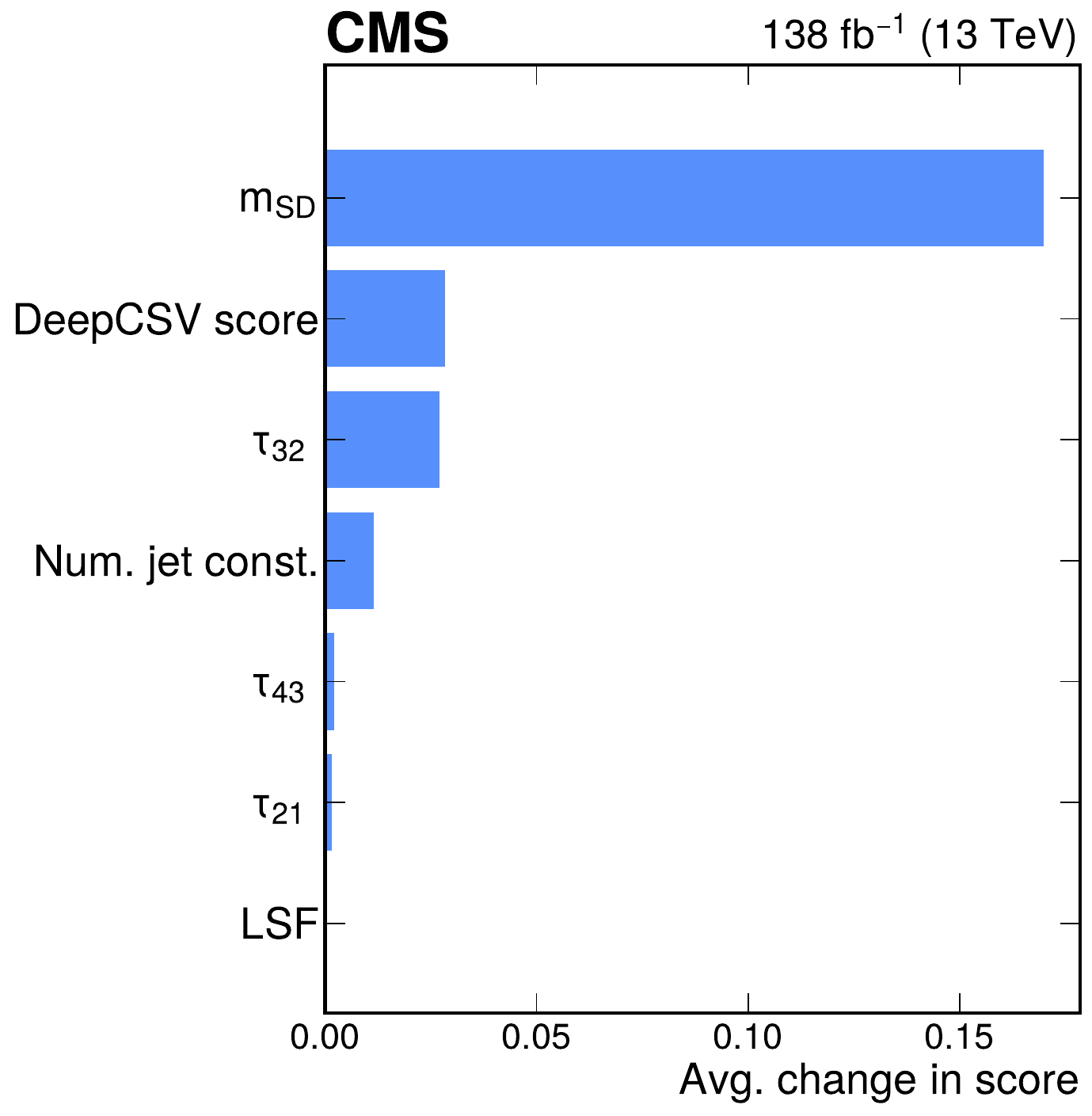}
    \caption{Post-fit jet mass in the fail (left) and pass (center) regions of the weakly supervised
    anomaly tag, for the 145 to 250\,GeV window: the $\mathrm{t\bar{t}}$ contribution (red) is
    recovered without labels. Right: permutation feature importance for the same model~\cite{MLG23002}.}
    \label{fig:top}
\end{figure}

A model-agnostic method is only convincing if it can find something known to be there. The weakly
supervised procedure was therefore applied to a lower-mass jet sample in which the "signal" is the SM
top quark, using only data and no truth labels~\cite{MLG23002}. In a jet mass window of 145 to
250\,GeV the classifier isolates a clear $\mathrm{t\bar{t}}$ contribution in the pass region that is
absent in the fail region, as shown in Fig.~\ref{fig:top}, with the peak near 175\,GeV from fully
merged top quark decays and the one near 80\,GeV from jets containing only the W boson. The naive
significance exceeds ten $\sigma$.

The same study answers the question of what one would do with a genuine excess. A permutation feature
importance analysis, which measures how much the anomaly score changes when one input observable is
randomized, identifies the soft-drop jet mass, the b tagging score, and the $N$-subjettiness ratio
$\tau_{32}$ as the three most important inputs. Comparing the distributions of these observables for
the most anomalous events with those of all events then reveals that the excess sits at a jet mass of
about 173\,GeV, contains a b quark, and has three prongs. The properties of the top quark are thus
reconstructed from the anomaly alone, which is exactly the chain of reasoning needed to characterize
a real discovery.

\section{Real-time anomaly detection in the level-1 trigger}
\label{sec:trigger}

An event rejected online can never be recovered, so a trigger built entirely around anticipated
signatures limits the reach of every downstream analysis. CMS has therefore deployed two unsupervised
anomaly detection algorithms directly in the level-1 (L1) trigger, which is implemented on
field-programmable gate arrays (FPGAs) and must decide at the full 40\,MHz LHC collision rate within
a latency budget of a few microseconds~\cite{MLG25001,DP2023079}.

AXOL1TL runs in the global trigger on the reconstructed objects available there, such as candidate
muons, jets, and global energy sums. It is a variational autoencoder, but at inference time only the
encoder is evaluated. CICADA runs earlier in the chain, in the
calorimeter trigger, and scores calorimeter energy deposits before any object reconstruction, giving
it access to information that object-based algorithms never see. It uses a convolutional autoencoder
as a teacher network and a much smaller student network, trained by knowledge distillation, to
predict the score directly. Both are trained on zero bias data, so no signal hypothesis enters at any
stage.

\begin{figure}[htbp]
    \centering
    \includegraphics[width=0.58\textwidth]{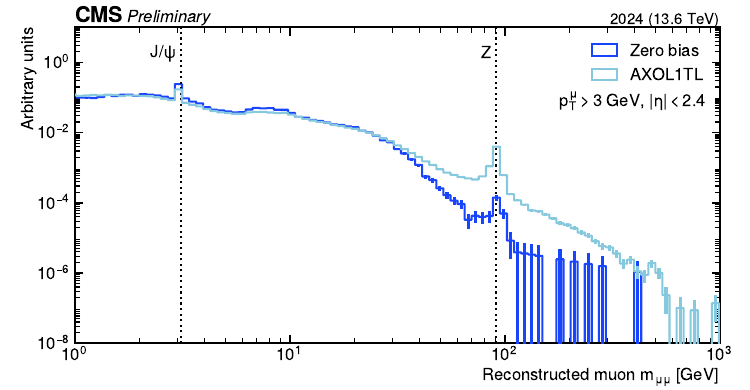}
    \caption{Dimuon invariant mass spectrum for events selected by the AXOL1TL anomaly trigger~\cite{MLG25001}.}
    \label{fig:axo}
\end{figure}

The algorithms operated throughout 2024 data taking at $\sqrt{s} = 13.6$\,TeV and selected more than
four billion events for permanent storage, at a cost of less than one percent of the total L1 rate in
uniquely collected events. That the selected data are physically sensible is shown in
Fig.~\ref{fig:axo}: well-known SM resonances, including the $\mathrm{J}/\psi$ and the Z boson, stand
out clearly in the dimuon mass spectrum of AXOL1TL events, and the Z boson, being rarer and more
energetic, is enhanced more strongly. This is not a search, but it establishes that an algorithm
given no model information concentrates on exactly the events a physicist would consider
interesting. The next step is to search this data set for signatures that conventional triggers
cannot select.

\vspace{-20pt}
\section{Toward event-level anomaly detection}
\label{sec:eventlevel}

All of the searches above locate the anomaly inside a jet, and a natural generalization is to let it
live in the event as a whole. Reference~\cite{EventLevel} introduces the di-object plus X topology
for weakly supervised anomaly detection: a resonance decaying to two SM particles provides the mass
variable in which a bump is sought, while the remaining activity X is what the classifier is trained
on, described by physically motivated variables derived from the geometry of the phase space manifold
of the collision. On benchmark signals the approach reaches discovery-level significance for models
that a conventional bump hunt in the same final state would miss. Within CMS, this strategy is now
being pursued in a dimuon plus X final state using the scouting data stream.

\section{Summary}
\label{sec:summary}

Model-independent searches based on anomaly detection have moved from proposal to practice at CMS.
Offline, five independent methods have been validated on Run 2 data, shown to be more sensitive than
an inclusive dijet search across a wide range of substructure signatures while remaining below the
reach of dedicated searches, and applied to 138\,fb$^{-1}$ without finding a significant excess. They
have been demonstrated on data by recovering the top quark and its properties with no labels, which
shows that a real excess could be characterized rather than merely observed. In real time, two
unsupervised algorithms now run on FPGAs in the level-1 trigger and have already collected billions
of events enriched in rare processes. The next steps are event-level anomaly detection, deeper use of
the trigger-level and scouting data sets, and signatures beyond jets. Taken together, these
developments mark a shift in emphasis from looking for a particular model to looking for anything
interesting.

{\footnotesize
\setlength{\bibsep}{2pt plus 0.5pt}

}

\end{document}